\documentclass[sigconf,authordraft=false, anonymous=false]{acmart}
\AtBeginDocument{%
  \providecommand\BibTeX{{%
    \normalfont B\kern-0.5em{\scshape i\kern-0.25em b}\kern-0.8em\TeX}}}

\usepackage{xcolor}
\usepackage{color, colortbl}
\usepackage{soul}
\usepackage{hyperref}

\sethlcolor{black}
\makeatletter
\newif\if@blind
\@blindtrue 
\if@blind \sethlcolor{black}\else
   
\fi

\definecolor{Gray}{gray}{0.9}
\newcolumntype{g}{>{\columncolor{Gray}}c}

\usepackage{graphicx}
\usepackage{multirow}

\usepackage{url}

\usepackage{subfig}
\usepackage{float}

\newcommand*{\origrightarrow}{}

\renewcommand*{\textrightarrow}{\fontfamily{cmr}\selectfont\origrightarrow}

\begin{document}

\title{Detection, Attribution and Localization of GAN Generated Images}
%

\author{Michael Goebel}
\email{mgoebel@ucsb.edu}
\affiliation{%
  \institution{University of California, Santa Barbara}
  \city{Santa Barbara}
  \state{California}
}

\author{Lakshmanan Nataraj}
\email{nataraj@mayachitra.com}
\affiliation{%
  \institution{Mayachitra, Inc.}
  \city{Santa Barbara}
  \state{California}
}

\author{Tejaswi Nanjundaswamy}
\email{tejaswi@mayachitra.com}
\affiliation{%
  \institution{Mayachitra, Inc.}
  \city{Santa Barbara}
  \state{California}
}

\author{Tajuddin Manhar Mohammed}
\email{mohammed@mayachitra.com}
\affiliation{%
  \institution{Mayachitra, Inc.}
  \city{Santa Barbara}
  \state{California}
}

\author{Shivkumar Chandrasekaran}
\email{shiv@mayachitra.com}
\affiliation{%
  \institution{Mayachitra, Inc.}
}
\affiliation{%
  \institution{University of California}
  \city{Santa Barbara}
  \state{California}
}

\author{B.S. Manjunath}
\email{manj@mayachitra.com}
\affiliation{%
  \institution{Mayachitra, Inc.}
}
\affiliation{%
  \institution{University of California}
  \city{Santa Barbara}
  \state{California}
}

\renewcommand{\shortauthors}{Goebel, et al.}

\begin{abstract}

Recent advances in Generative Adversarial Networks (GANs) have led to the creation of realistic-looking digital images that pose a major challenge to their detection by humans or computers. 
GANs are used in a wide range of tasks, from modifying small attributes of an image (StarGAN~\cite{choi2018stargan}), transferring attributes between image pairs (CycleGAN~\cite{zhu2017unpaired}), as well as generating entirely new images (ProGAN~\cite{karras2017progressive}, StyleGAN~\cite{karras2019style}, SPADE/GauGAN~\cite{park2019semantic}).
In this paper, we propose a novel approach to detect, attribute and localize GAN generated images that combines image features with deep learning methods.
For every image, co-occurrence matrices are computed on neighborhood pixels of RGB channels in different directions (horizontal, vertical and diagonal).
A deep learning network is then trained on these features to detect, attribute and localize these GAN generated/manipulated images.
A large scale evaluation of our approach on 5 GAN datasets comprising over 2.76 million images (ProGAN, StarGAN, CycleGAN, StyleGAN and SPADE/GauGAN) shows  promising results in detecting GAN generated images. 

\end{abstract}


\maketitle

\section{Introduction}
\label{s:intro}

{T}{he} advent of Convolutional Neural Networks (CNNs)~\cite{krizhevsky2012imagenet,simonyan2014very} has shown application in a wide variety of image processing tasks, and image manipulation is no exception. 
In particular, Generative Adversarial Networks (GANs)~\cite{goodfellow2014generative} have been one of the most promising advancements in image enhancement and manipulation - the generative Artificial Intelligence (AI) patents grew by 500\% in 2019~\cite{patent-gans}.
Due to the success of using GANs for image editing, it is now possible to use a combination of GANs and off-the-shelf image-editing tools to modify digital images to such an extent that it has become difficult to distinguish doctored images from normal ones.
In December 2019, Facebook announced that it removed hundreds of accounts whose profile pictures were generated using AI~\cite{ai-fb-remove1,ai-fb-remove2}.

\begin{figure}[t]
\centering
\includegraphics[width=0.45 \textwidth]{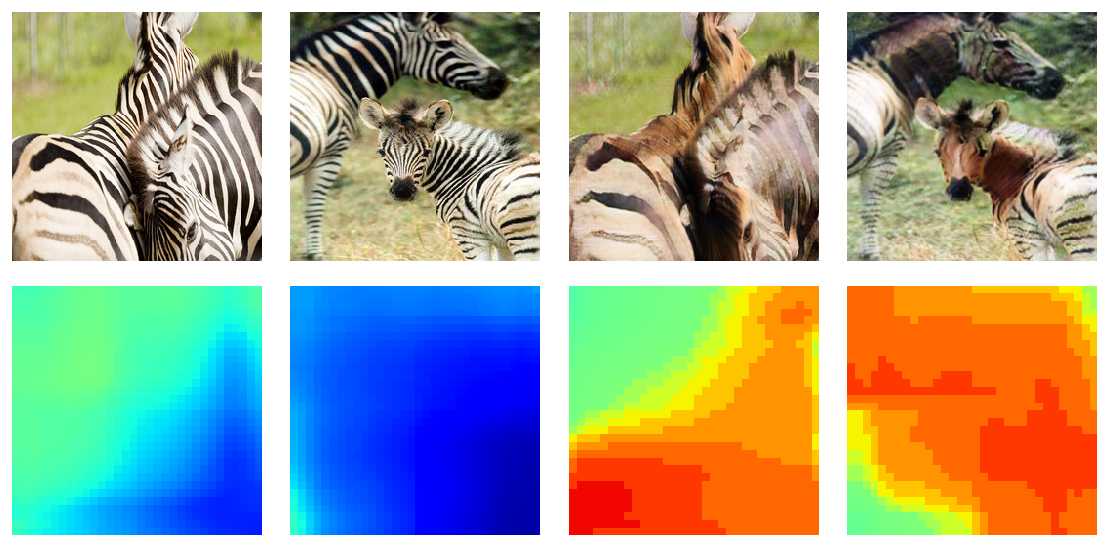}
\caption{Input test set images on the top row, and our proposed detection heatmaps on the bottom. The two images on the left are authentic zebra images, those on the right are generated using CycleGAN.}
\label{fig:gan-eg-z2h}
\end{figure}

The GAN training procedure involves a generator and discriminator. 
The generator may take in an input image and a desired attribute to change, then output an image containing that attribute. 
The discriminator will then try to differentiate between images produced by the generator and the authentic training examples.
The generator and discriminator are trained in an alternate fashion, each attempting to optimize its performance against the other. 
Ideally, the generator will converge to a point where the output images are so similar to the ground truth that a human will not be able to distinguish the two.
In this way, GANs have been used to produce ``fake'' images that are very close to the real input images.
These include image-to-image attribute transfer (CycleGAN~\cite{zhu2017unpaired}), generation of facial attributes and expressions (StarGAN~\cite{choi2018stargan}), as well as generation of whole new images such as faces (ProGAN~\cite{karras2017progressive}, StyleGAN~\cite{karras2019style}), indoors (StyleGAN) and landscapes (SPADE/GauGAN~\cite{park2019semantic}). 
In digital image forensics, the objective is to both detect these fake GAN generated images, localize areas in an image which have been generated by GANs,  as well as identify which type of GAN was used in generating the fake image.


 



In the GAN training setup, the discriminator functions directly as a classifier of GAN and non-GAN images.
So the question could be raised as to \emph{why not use the GAN discriminator to detect if it's real or fake?}
To investigate this, we performed a quick test using the CycleGAN algorithm under the maps-to-satellite-images category, where fake maps are generated from real satellite images, and vice versa.
In our test, we observed that the discriminator accuracy over the last 50 epochs was only 80.4\%. 
However, state-of-the-art deep learning detectors for CycleGAN often achieve over 99\% when tested on the same type of data which they are trained~\cite{marra2018detection,nataraj2019detecting,zhang2019detecting}. 
Though the discriminator fills its role of producing a good generator, it does not compare performance wise to other methods which have been suggested for detection.



While the visual results generated by GANs are promising, the GAN based techniques alter the statistics of pixels in the images that they generate.  
Hence, methods that look for deviations from natural image statistics could be effective in detecting GAN generated fake images.
These methods have been well studied in the field of steganalysis which aims to detect the presence of hidden data in digital images.
One such method is based on analyzing co-occurrences of pixels by computing a co-occurrence matrix.
Traditionally, this method uses hand crafted features computed on the co-occurrence matrix and a machine learning classifier such as support vector machines determines if a message is hidden in the image~\cite{sullivan2005steganalysis,sullivan2006steganalysis}.
Other techniques involve calculating image residuals or passing the image through different filters before computing the co-occurrence matrix~\cite{pevny2010steganalysis,fridrich2012rich, cozzolino2017recasting}.
Inspired by steganalysis and natural image statistics, we propose a novel method to identify GAN generated images using a combination of pixel co-occurrence matrices and deep learning.
Here we pass the co-occurrence matrices directly through a deep learning framework and allow the network to learn important features of the co-occurrence matrices.
This also makes it difficult to perform adversarial perturbations on the co-occurrence matrices since the underlying statistics will be altered.
We also avoid computation of residuals or passing an image through various filters which results in loss of information.
We rather compute the co-occurrence matrices on the image pixels itself.
For detection, we consider a two class framework - real and GAN, where a network is trained on co-occurrence matrices computed on the whole image to detect if an image is real or GAN generated. 
For attribution, the same network is trained in a multi-class setting depending on which GAN the image was generated from. 
For localization, a network is trained on co-occurrence matrices computed on image patches and a heatmap was is generated to indicate which patches are GAN generated.
Detailed experimental results on large scale GAN datasets comprising over 2.76 million images originating from multiple diverse and challenging datasets generated using GAN based methods show that our approach is promising and will be an effective method for tackling future challenges of GANs.


The main contributions of the paper are as follows:

\begin{itemize}
    \item We propose a new method for detection, attribution and localization of GAN images using a combination of deep learning and co-occurrence matrices.
    \item We compute co-occurrence matrices on different directions of an image and then train them using deep learning. For detection and attribution, the matrices are computed on the whole image and for localization, the matrices are computed on image patches to obtain a heatmap.
    \item We perform our tests on over 2.7 million images, which to our knowledge, is the largest evaluation on detection of GAN images.
    \item We provide explainability of our approach using t-SNE visualizations on different GAN datasets.
    \item We show the method holds under both varying JPEG compression factors and image patch sizes, accommodating a range of real-world use cases.
\end{itemize}

\section{Related Work}
\label{s:rw}

Since the seminal work on GANs~\cite{goodfellow2014generative}, there have been several hundreds of papers on using GANs to generate images. 
These works focus on generating images of high perceptual quality~\cite{mirza2014conditional, radford2015unsupervised,salimans2016improved,isola2017image,arjovsky2017wasserstein,gulrajani2017improved, karras2017progressive}, image-to-image translations~\cite{isola2017image,yi2017dualgan,zhu2017unpaired}, domain transfer~\cite{taigman2016unsupervised,kim2017learning}, super-resolution~\cite{ledig2017photo}, image synthesis and completion~\cite{li2017generative,iizuka2017globally,wang2018high}, and generation of facial attributes and expressions~\cite{liu2016coupled, perarnau2016invertible,kim2017learning,choi2018stargan}.
Several methods have been proposed in the area of image forensics over the past years~\cite{forgery_1,forgery_2,forgery_3,forgery_4,forgery_5}. 
Recent approaches have focused on applying deep learning based methods to detect tampered images~\cite{bayar2016deep, bayar2017design,rao2016deep,bunk2017detection,bappy2017exploiting,cozzolino2017recasting,zhou2018learning}.

In digital image forensics, detection of GAN generated images has been an active topic in recent times and several papers have been published in the last few years~\cite{marra2018detection,valle2018tequilagan,li2018detection,mccloskey2018detecting,li2018can,jain2018on,tariq2018detecting,marra2018gans,mo2018fake,do2018forensics,yu2018attributing,hsu2018learning,nataraj2019detecting,zhang2019detecting,wang2019fakespotter,marra2019incremental,albright2019source,zhuang2019detecting,he2019detection,neves2019real,zhang2019no,wang2019cnn,kim2019stylegan,jain2020detecting,bonettini2020use,frank2020leveraging,guo2020fake,chen2020manipulated,bonettini2020use}.
Other similar research include detection of computer generated (CG) images~\cite{dirik2007new,wu2011identifying,mader2017identifying,rahmouni2017distinguishing}

In~\cite{marra2018detection}, Marra et al. compare various methods to identify CycleGAN images from normal ones.
The top results they obtained are using a combination of residual features~\cite{cozzolino2014image,cozzolino2017recasting} and deep learning~\cite{chollet2017xception}.
In~\cite{li2018detection}, Li et al. compute the residuals of high pass filtered images and then extract co-occurrence matrices on these residuals, which are then concatenated to form a feature vector that can distinguish real from fake GAN images.
In~\cite{zhang2019detecting}, Zhang et al. identify an artifact caused by the up-sampling component
included in the common GAN pipeline and show that such artifacts are manifested as replications of spectra in the
frequency domain and thus propose a classifier model based on the spectrum input, rather than the pixel input.

We had previously proposed a 3 channel co-occurrence matrix based method~\cite{nataraj2019detecting}, and many other papers have shown the efficacy of this method in their experimental evaluations \cite{liu2020global,neves2019ganprintr,tolosana2020deepfakes,neves2019real,zhang2019no,hulzebosch2020detecting,mansourifar2020one}.
However, in this paper we compute co-occurrence matrices on horizontal, vertical and diagonal directions, as well as compute them on image patches, thus facilitating detection, attribution and localization of GAN generated images. 

\section{Methodology}

\subsection{Co-Occurrence Matrix Computation}
The co-occurrence matrices represent a two-dimensional histogram of pixel pair values in a region of interest. The vertical axis of the histogram represents the first value of the pair, and the horizontal axis, the second value. Equation \ref{eq:co_mtx_comp} shows an example of this computation for a vertical pair.

\begin{equation}
    C_{i,j} = \sum_{m,n} \begin{cases}
    1, & I[m,n] = i \medspace and \medspace I[m+1,n] = j \\
    0, & otherwise
    \end{cases}
    \label{eq:co_mtx_comp}
\end{equation}

Under the assumption of 8-bit pixel depth, this will always produce a co-occurrence matrix of size 256x256. This is a key advantage of such a method, as it will allow for the same network to be trained and tested on a variety of images without resizing.

Which pairs of pixels to take was one parameter of interest in our tests. For any pixel not touching an edge, there are 8 possible neighbors. We consider only 4 of these for our tests; right, bottom right, bottom, and bottom left. The other 4 possible pairs will provide redundant information. For example, the left pairs are equivalent to swapping the order of the first and second pixel in the right pair. In the co-occurrence matrix, this corresponds to a simple transpose. There are many subsets of these 4 pairs which could be taken, but our tests consider only a few; horizontal, vertical, horizontal and vertical, or all.

\begin{figure}[t]
\centering
\captionsetup{justification=centering}
\includegraphics[scale=0.25]{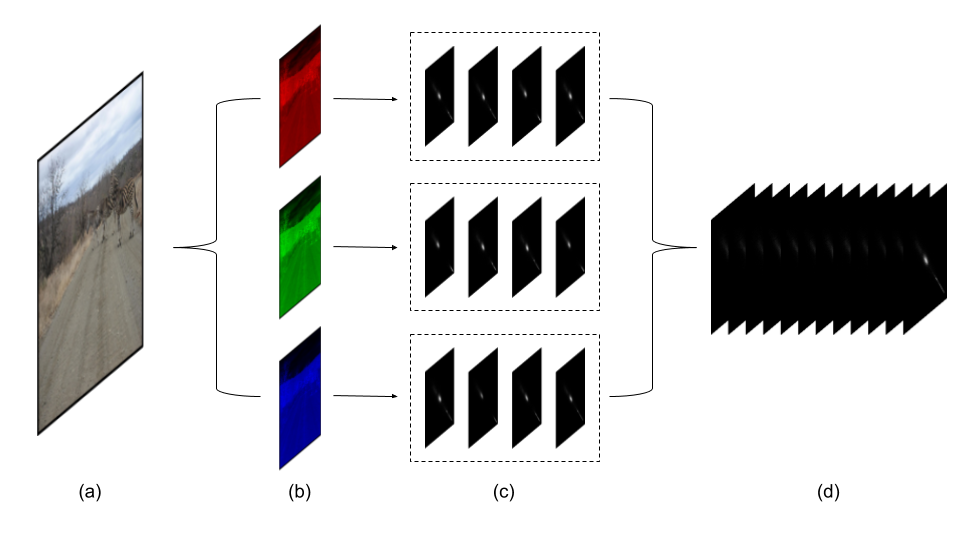}
\vspace{-5pt}
\caption{An example co-occurrence computation. The input image (a) is split into its three color channels (b). For each color channel, 4 different pairs of pixels are used to generate 2-dimensional histograms (c). Horizontal, vertical, diagonal, and anti-diagonal pairs are considered. These histograms are then stacked to produce a single tensor (d). For some tests, only a subset of the co-occurrence matrices will be used.}
\label{fig:co_mtx_diagram}
\end{figure}

Before passing these matrices through a CNN, some pre-processing is done. First, each co-occurrence matrix is divided by its maximum value. Given that the input images may be of varying sizes, this will force all inputs into a consistent scale. After normalization, all co-occurrence matrices for an image are stacked in the depth dimension. In the example of an RGB image with all 4 co-occurrence pairs, this will produce a new image-like feature tensor of size 256x256x12. Figure \ref{fig:co_mtx_diagram} gives a visualization of this process.

\subsection{Convolutional Neural Networks} 
\label{cnn_section}
While the co-occurrence matrices are not themselves images, treating them as so has some theoretical backing. One of the primary motivations for using CNNs in image processing is their translation invariance property. In the case of a co-occurrence matrix, a translation along the main diagonal corresponds to adding a constant value to the image. 
We would not expect this manipulation to affect the forensic properties.

\begin{figure}
    \centering
    \includegraphics[width=0.5\textwidth]{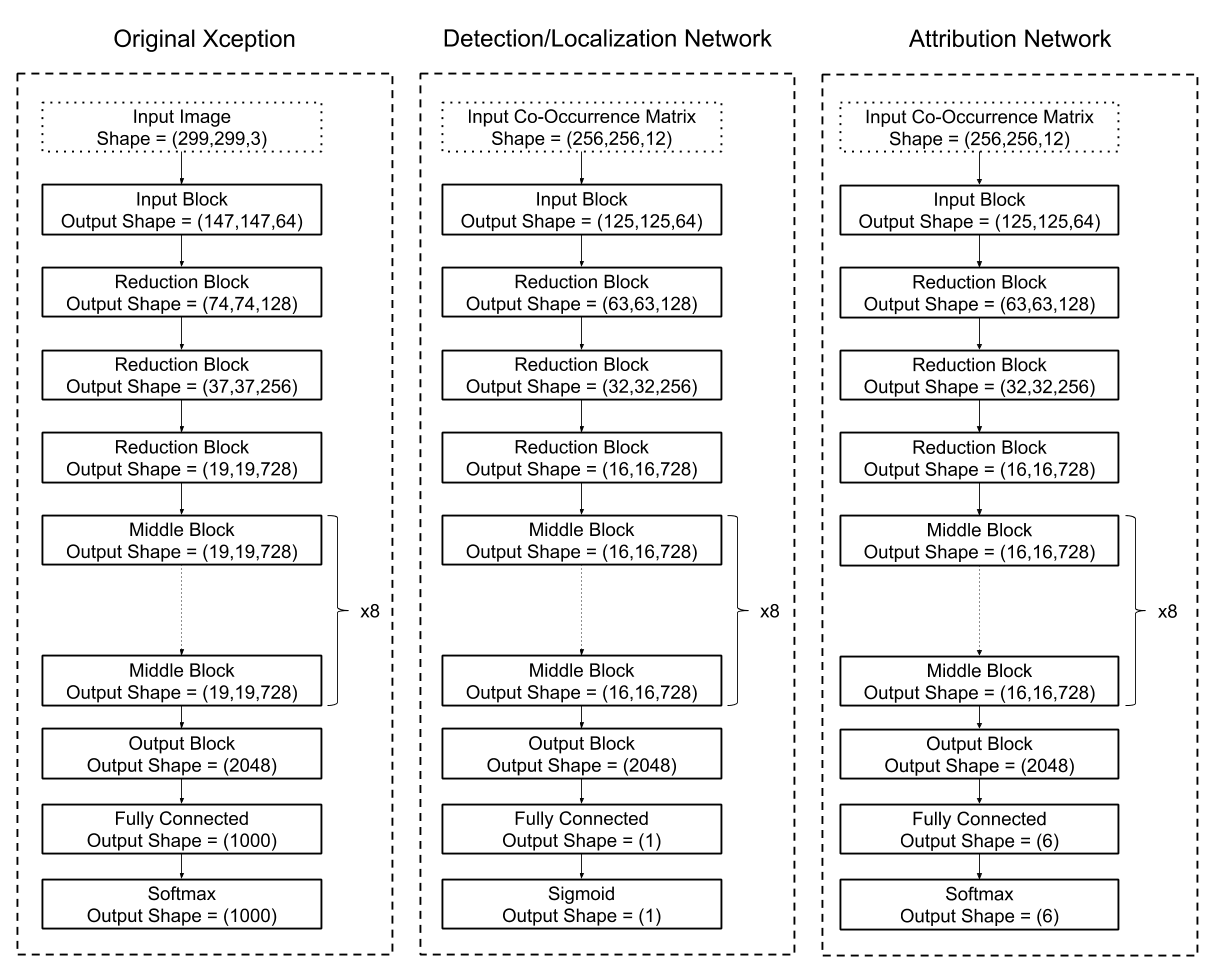}
    \caption{The original Xception network~\cite{chollet2017xception}, shown next to our two modified models. Our architectures for detection and attribution are the same, except for the last layer and activation.}
    \label{fig:our_xception}
\end{figure}

In this paper, we use Xception Net~\cite{chollet2017xception} deep neural network architecture for detection, attribution and localization of GAN generated images.
The Xception network is a modified version of Inception network~\cite{szegedy2017inception} but  was created under a stronger theoretical assumption than the original Inception, where cross-channel correlations are completely split from spatial correlations by use depth-wise separable convolutions. 
The network also includes residual connections, as shown in Figure~\ref{fig:our_xception}.
For these reasons, the authors claim that Xception can more easily find a better convergence point than most other CNN architectures, while keeping model capacity low~\cite{chollet2017xception}.
In this paper, we modify the original input and output shapes in the Xception network to accommodate our task as shown in Figure~\ref{fig:our_xception}. 
The initial convolutional portions of the network remain unchanged, though the output sizes of each block are slightly different. 
This small change in size is accommodated by the global pooling step. 
Finally, the last fully connected layer of each network is changed to the desired number of output classes, and given the appropriate activation.
For detection and attribution, our architectures are the same except for the last layer and activation.
For localization, no changes were made to the model architecture but co-occurrence matrices were extracted on small image patches, and individually passed through the network.



\section{Datasets}
\begin{figure*}[t]
\centering
\subfloat[Real Images]{\includegraphics[width=0.48\textwidth]{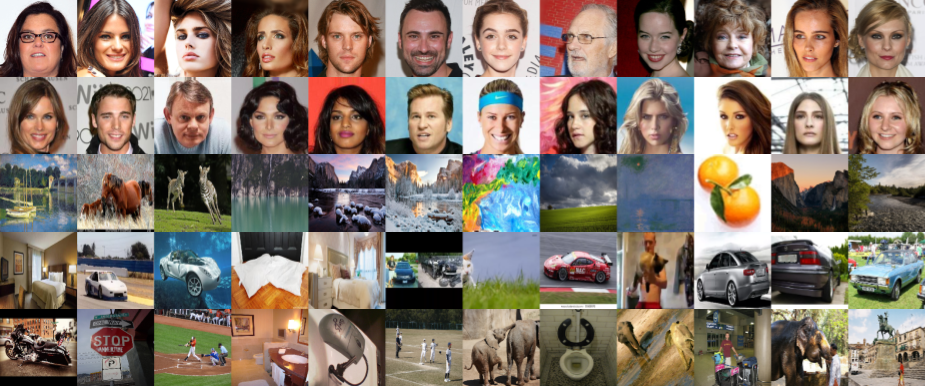}} \ \ 
\subfloat[GAN Images]{\includegraphics[width=0.48\textwidth]{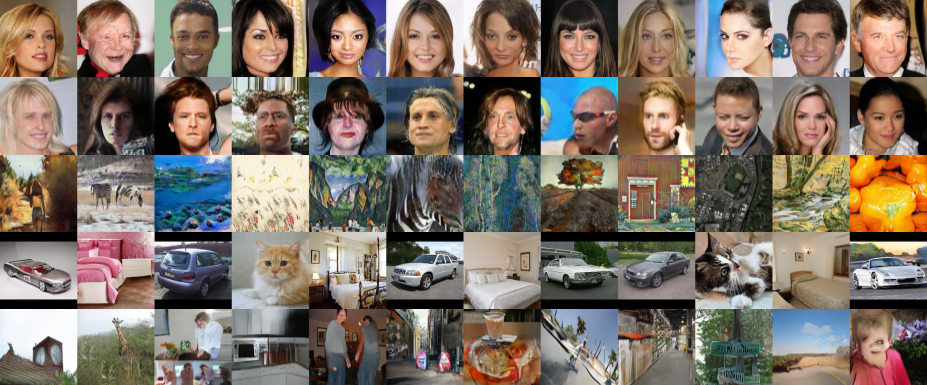}}
\caption{Sample images from different GAN datasets (a) Real images and (b) GAN images from different GAN datasets (top to bottom): ProGAN~\cite{karras2017progressive}, StarGAN~\cite{choi2018stargan}, CycleGAN~\cite{zhu2017unpaired}, StyleGAN~\cite{karras2019style}, and SPADE/GauGAN~\cite{park2019semantic}.}
\label{fig:gan_datasets_samples}
\end{figure*}

\begin{figure}
    \centering
    \includegraphics[width=0.4\textwidth]{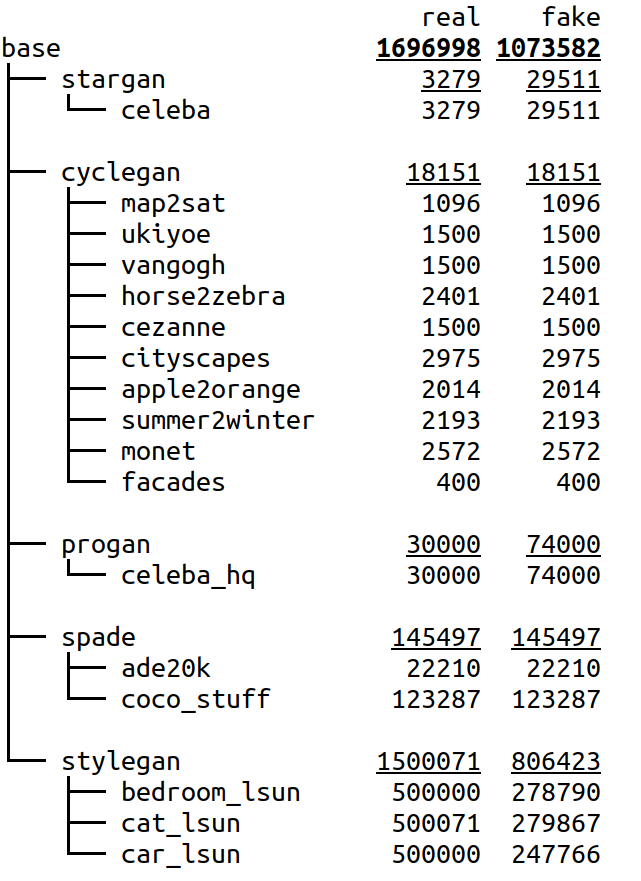}
    \caption{Quantitative summary of the GAN datasets used in our experiments.}
    \label{fig:gan_dataset_sum}
\end{figure}

We evaluated our method on five different GAN architectures, of which each was trained on several different image generation tasks: ProGAN~\cite{karras2017progressive}, StarGAN~\cite{choi2018stargan}, CycleGAN~\cite{zhu2017unpaired}, StyleGAN~\cite{karras2019style}, and SPADE/GauGAN~\cite{park2019semantic}. 
The modifications included image-to-image translation, facial attribute modification, style transfer, and pixel-wise semantic label to image generation.
A summary of the datasets, including the number of images from each class, is shown in Figure~\ref{fig:gan_dataset_sum}.
These comprise a total of more than 2.76 million images of which 1.69 million images are real images and 1.07 million images are fake GAN generated images.
In several cases, one or more images in the GAN generated category will be directly associated with an image in the authentic class. 
For example, a person's headshot untampered, blond, aged, and gender reversed will all be in the dataset. 
However, the splitting for training accounts for this, and will keep all of these images together to be put into either training, validation, or test.
Some sample images from all the GAN datasets are shown in Figure~\ref{fig:gan_datasets_samples}. 

\subsubsection{StarGAN}
This dataset consists of only celebrity photographs from the CelebA dataset~\cite{liu2015deep}, and their GAN generated counterparts~\cite{choi2018stargan}. The GAN changes attributes of the person to give them black hair, brown hair, blond hair, different gender, different age, different hair and gender, different hair and age, different gender and age, or different hair, age, and gender. These are the smallest of all of the training images, being a square of size 128 pixels.

\subsubsection{CycleGAN}
This datasets includes image-to-image translations between a wide array of image classes~\cite{zhu2017unpaired}. The sets horse2zebra, apple2orange, and summer2winter do a strict image-to-image translation, with the assumption that the GAN will learn the areas to modify. While the whole output is generated by the GAN, the changes for these will ideally be more localized. Ukiyoe, Vangogh, Cezanne, and Monet are four artists which the GAN attempts to learn a translation from photographs to their respective styles of painting. Facades and cityscapes represent the reverse of the image segmentation task. Given a segmentation map as input, they produce an image of a facade or cityscape. Map2sat takes in a Google Maps image containing road, building, and water outlines, and generates a hypothetical satellite image.

\subsubsection{ProGAN}
This dataset consists of images of celebrities, and their GAN generated counterparts, at a square size of 1024 pixels~\cite{karras2017progressive}. All data was obtained per the instructions provided in the paper's Github repository.

\subsubsection{SPADE/GauGAN}
SPADE/GauGAN contains realistic natural images generated using GANs~\cite{park2019semantic}.
This dataset uses images from ADE20k~\cite{zhou2017scene} dataset containing natural scenes and COCO-Stuff~\cite{caesar2018coco} dataset comprising day-to-day images of things and other stuff, along with their associated segmentation maps. 
These untampered images are considered as real images in the GAN framework, and the pretrained models provided by the SPADE/GauGAN authors are used to generate GAN images from the segmentation maps.

\subsubsection{StyleGAN}
This dataset contains realistic images of persons, cars, cats and indoor scenes~\cite{karras2019style}.
Images for this dataset were provided by the authors. 

\section{Experiments}

\subsection{Training Procedure}
All deep learning experiments in this paper were done using Keras 2.2.5 and all training was done using an Adam optimizer~\cite{kingma2014adam}, learning rate of $10^{-4}$, and cross-entropy loss.
A batch size of 64 was used for all experiments. Unless otherwise stated, a split of 90\% training, 5\% validation, and 5\% test was used.
Given the large amount of data available, a single iteration through the entire dataset for training took 10 hours on a single Titan RTX GPU.
To allow for more frequent evaluation on the validation set, the length of an epoch was capped at 100 batches. Validation steps were also capped at 50 batches, and test sets at 2000 batches.
After training for a sufficient period of time for the network to converge, the checkpoint which scored the highest in validation was chosen for testing. For experiments to determine hyper-parameters, training was capped at 50 epochs, and took approximately 3 hours each on a single Titan RTX. 
After determination of hyper-parameters, training of the final model was done for 200 epochs, taking approximately 12 hours.


\subsection{Comparison with other CNN architectures:}
First we evaluate our method on different well known CNN architectures: VGG16~\cite{simonyan2014very}, ResNet50 and ResNet101~\cite{he2016deep},ResNet50V2, ResNet101V2 and ResNet152V2~\cite{he2016identity}, InceptionV3 and InceptionResNetV2~\cite{szegedy2017inception}, and Xception~\cite{chollet2017xception}. 
Shown in Table~\ref{tab:cnn_arch} are the results for the different CNN networks. 
Though designed for ImageNet classification, all models take in an image with height, width, and 3 channels, and output a one-hot encoded label. 
The models are used as-is, with the following slight modifications. 
First, the number of input channels is set to be the depth of the co-occurrence feature tensor.
Second, input shape was fixed at 256x256.
Third, the number of output channels was set to 1. 
All of these parameters were passed as arguments to the respective Keras call for each model. 
A small margin separated the top performers, though Xception was the best with an accuracy of \textbf{0.9916} and had fewer parameters than others. 
For this reason, we chose Xception for the remainder of the experiments. 


\begin{table}
\centering
\caption{Comparison of different popular ImageNet~\cite{imagenet_cvpr09} classification architectures on classifying GANs from co-occurrence matrices. All datasets are mixed for training, validation, and testing. The features are extracted from a whole image, with no JPEG compression.}
\begin{tabular}{|c|c|}
\hline
Network & Accuracy \\
\hline
VGG16~\cite{simonyan2014very} & 0.6115 \\
\hline
ResNet50~\cite{he2016deep} & 0.9677 \\
\hline
ResNet101~\cite{he2016deep} & 0.9755 \\
\hline
ResNet152V2~\cite{he2016identity} & 0.9795 \\
\hline
ResNet50V2~\cite{he2016identity} & 0.9856 \\
\hline
InceptionResNetV2~\cite{szegedy2017inception} & 0.9885 \\
\hline
InceptionV3~\cite{szegedy2016rethinking} & 0.9894 \\
\hline
ResNet101V2~\cite{he2016identity} & 0.9900 \\
\hline
Xception~\cite{chollet2017xception} & \textbf{0.9916} \\
\hline
\end{tabular}
\label{tab:cnn_arch}
\end{table}

\subsection{Comparison of Co-occurrence Matrix Pairs}
Next we perform tests with different co-occurrence pairs, shown in Table~\ref{tab:diff_pairs}. 
These experiments included JPEG compression, randomly selected from quality factors of 75, 85, 90, and no compression. 
Interestingly, it seems that the addition of more co-occurrence pairs did not significantly improve performance. 
For the remainder of the test, all 4 co-occurrence pairs were used.

\begin{table}
\centering
\caption{Test on difference co-occurrence pairs. These were done on the whole image, with the additional challenge of JPEG compression. The JPEG quality factor was randomly selected with equal probability from the set of 75, 85, 90, or no JPEG compression}
\begin{tabular}{|c|c|}
\hline
Pairs & Accuracy \\
\hline
Horizontal & 95.51 \\
\hline
Vertical & 95.56 \\
\hline
Hor and Ver & 95.17 \\
\hline
Hor, Ver, and Diag & \textbf{95.68}\\
\hline
\end{tabular}
\label{tab:diff_pairs}
\end{table}

\subsection{Effect of patch size}
For applications, the two parameters of interest were JPEG compression and patch size. The results for different patch sizes are shown in Table~\ref{tab:diff_patch}. These results are from images JPEG compressed by a factor randomly selected from 75, 85, 90, and none. A model is trained for each of the possible patch sizes, and then each model is tested against features from each patch size. It should be noted that in cases where the input image is smaller than the requested patch size, the whole image is used.
There is notable generalization between different patch sizes, in that the model trained on a patch size of 256 and tested on 128 achieves an accuracy within a few percentage points of a model trained and tested on 128. Thus we would expect our models to work with a variety of untested patch sizes within a reasonable range while only taking a minor performance drop.

\begin{table}
\centering
\caption{Accuracy when trained on one patch size, and tested on another. Data for training and testing has been pre-processed using JPEG compression with quality factors randomly selected from 75, 85, 90 or none.}
\begin{tabular}{|c|c|c|c|c|}
\hline
\multicolumn{2}{|c|}{\multirow{2}{*}{}} & \multicolumn{3}{c|}{Train} \\
\cline{3-5}
\multicolumn{2}{|c|}{} & 64 & 128 & 256 \\
\hline
\multirow{3}{*}{Test} & 64 & 0.7814 & 0.7555 & 0.6778 \\
\cline{2-5}
 & 128 & 0.8273 & 0.8336 & 0.8158 \\
\cline{2-5}
 & 256 & 0.8311 & 0.8546 & 0.8922 \\
\hline
\end{tabular}
\label{tab:diff_patch}
\end{table}

\subsection{Effect of JPEG compression}
Now assuming a fixed patch size of 128, we varied the JPEG quality factors: 75,85,90 and no compression.
The model was again trained only on one particular JPEG factor as shown in Table~\ref{tab:diff_jpeg}. As expected, we see that performance increases with respect to quality factor. However, this table also shows that the model does not overfit to a particular quality factor, in that testing on a slightly better or worse quality factor gives a score not far from a model tuned to the particular test quality factor.

\begin{table}
\centering
\caption{Test accuracy when model is trained on images pre-processed with one JPEG quality factor, and tested on another.}
\begin{tabular}{|c|c|c|c|c|c|}
\hline
\multicolumn{2}{|c|}{\multirow{2}{*}{}} & \multicolumn{4}{c|}{Train} \\
\cline{3-6}
\multicolumn{2}{|c|}{} & 75 & 85 & 90 & None \\
\hline
\multirow{4}{*}{Test} & 75 & 0.7738 & 0.7448 & 0.7101 & 0.6605 \\
\cline{2-6}
 & 85 & 0.8209 & 0.8593 & 0.8362 & 0.7209 \\
\cline{2-6}
 & 90 & 0.8310 & 0.8690 & 0.8756 & 0.7651 \\
\cline{2-6}
 & None & 0.9198 & 0.9386 & 0.9416 & 0.9702 \\
\hline
\end{tabular}
\label{tab:diff_jpeg}
\end{table}

\subsection{Generalization}
To test the generalization between GANs, leave-one-out cross validation was used for each GAN architecture. One dataset of GAN images is used for testing and remaining GAN image datasets are used for training.
Here, a patch size of 128 was used with no JPEG compression. 
From Table~\ref{tab:loo_gan}, we see that some GAN datasets such as SPADE, StarGAN and StyleGAN have high accuracy and are more generalizable. 
However, the accuracies for CycleGAN and ProGAN are lower in comparison, thus suggesting that images from these GAN categories should not be discarded when building a bigger GAN detection framework.
We also considered computing co-occurrence matrices on the whole image and then repeated the above experiment, but the overall accuracy did not improve.

\begin{table}
\centering
\caption{Train on all but one GAN, test on the held out images. Patch size of 128, no JPEG compression.}
\begin{tabular}{|c|c|}
\hline
Test GAN & Accuracy \\
\hline
StarGAN & 0.8490 \\
\hline
CycleGAN & 0.7411 \\
\hline
ProGAN & 0.6768 \\
\hline
SPADE & 0.9874\\
\hline
StyleGAN & 0.8265 \\ 
\hline
\end{tabular}
\vspace{-10pt}
\label{tab:loo_gan}
\end{table}

\begin{figure}[t]
    \centering
    \includegraphics[width=0.48\textwidth]{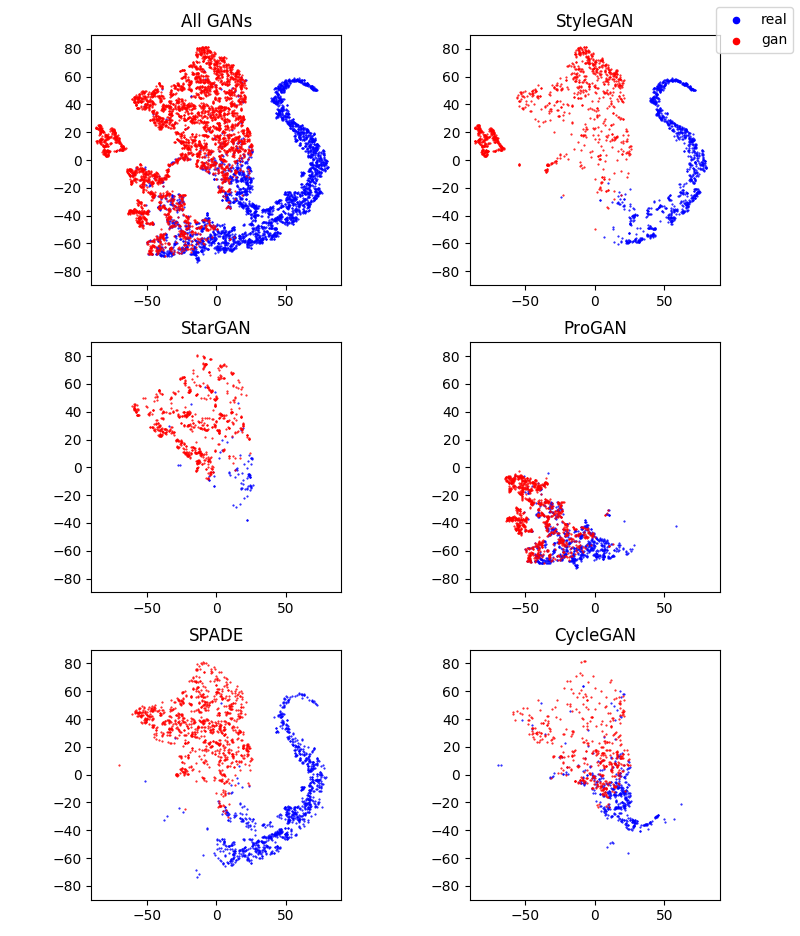}
    \vspace{-5pt}
    \caption{Visualization of images from different GAN datasets using t-SNE~\cite{maaten2008visualizing}.}
    \label{fig:tsne}
\end{figure}

\noindent \textit{Visualization using t-SNE:} To further investigate the variability in the GAN detection accuracies under the leave-one-out setting, we use t-SNE visualization~\cite{maaten2008visualizing} from outputs of the penultimate layer of the CNN, using images from the test set (as shown in Figure~\ref{fig:tsne}). 
The t-SNE algorithm aims to reduce dimensionality of a set of vectors while preserving relative distances as closely as possible.
With an L2 distance function and linear transformation, this can be efficiently found by PCA. 
While there are many solutions to this problem for different distance metrics and optimization methods, KL divergence on the Student-t distribution used in t-SNE has shown the most promising results on real-world data~\cite{maaten2008visualizing}.

To limit computation time, no more than 1000 images were used for a particular GAN from either the authentic or GAN classes. As recommended in the original t-SNE publication, the vector was first reduced using Principle Component Analysis (PCA). 
The original 2048 were reduced to 50 using PCA, and passed to the t-SNE algorithm.
As we see in Figure~\ref{fig:tsne}, the images in CycleGAN and ProGAN are more tightly clustered, thus making them difficult to distinguish between real and GAN generated images, while the images from StarGAN, SPADE and StyleGAN are more separable, thus resulting in higher accuracies in the leave-one-out experiment.

\begin{table*}
\centering
\scriptsize
\caption{Comparison with State-of-the-art.}
\begin{tabular}{|c|c|c|c|c|c|c|c|c|c|c|c|}
\hline
Method & ap2or & ho2zeb & wint2sum & citysc. & facades & map2sat & Ukiyoe & Van Gogh & Cezanne & Monet & Average \\
\hline
Steganalysis feat. & 0.9893 & 0.9844 & 0.6623 & 1.0000 & 0.9738 & 0.8809 & 0.9793 & 0.9973 & 0.9983 & 0.9852 & 0.9440\\
\hline
Cozzalino2017 & 0.9990 & 0.9998 & 0.6122 & 0.9992 & 0.9725 & 0.9959 & 1.0000 & 0.9993 & 1.0000 & 0.9916 & 0.9507\\
\hline
XceptionNet & 0.9591 & 0.9916 & 0.7674 & 1.0000 & 0.9856 & 0.7679 & 1.0000 & 0.9993 & 1.0000 & 0.9510 & 0.9449\\
\hline
Nataraj2019 & 0.9978 & 0.9975 & 0.9972 & 0.9200 & 0.8063 & 0.9751 & 0.9963 & 1.0000 & 0.9963 & 0.9916 & 0.9784\\
\hline
Zhang2019 & 0.9830 & 0.9840 & 0.9990 & 1.0000 & 1.0000 & 0.7860 & 0.9990 & 0.9750 & 0.9920 & 0.9970 & 0.9720\\
\hline
Proposed approach & 0.9982 & 0.9979 & 0.9982 & 0.9366 & 0.9498 & 0.9776 & 0.9973 & 0.9980 & 0.9993 & 0.9697 & \textbf{0.9817}\\
\hline
\end{tabular}
\label{tab:comp1}
\end{table*}

\subsection{Comparison with State-of-the-art}
\label{sec:comparison}

We compare our proposed approach with various state-of-the-art methods~\cite{marra2018detection,nataraj2019detecting,zhang2019detecting} on the CycleGAN dataset. 
In~\cite{marra2018detection}, Marra et al. proposed the leave-one-category-out benchmark test to see how well their methods work when one category from the CycleGAN dataset is kept for testing and remaining are kept for training. 
The methods they evaluated are based on steganalysis, generic image manipulations, detection of computer graphics, a GAN discriminator used in the CycleGAN paper, and generic deep learning architecture pretrained on ImageNet~\cite{imagenet_cvpr09}, but fine tuned to the CycleGAN dataset. 
Among these the top preforming ones were from steganalysis~\cite{fridrich2012rich,cozzolino2014image} based on extracting features from high-pass residual images, a deep neural network designed to extract residual features~\cite{cozzolino2017recasting} (Cozzolino2017) and XceptionNet~\cite{chollet2017xception} deep neural network trained on ImageNet but fine-tuned to this dataset. 
Apart from Marra et al.~\cite{marra2018detection}, we also compare our method with approaches including Nataraj et al. (Nataraj2019)~\cite{nataraj2019detecting}, which uses co-occurrence matrices computed in the horizontal direction, and Zhang et al.(Zhang2019)~\cite{zhang2019detecting}, which uses spectra of up-sampling artifacts used in the GAN generating procedure to classify GAN images. 


Table~\ref{tab:comp1} summarizes the results of our proposed approach against other state-of-the-art approaches. 
Our method obtained the best average accuracy of \textbf{0.9817}, when compared with other methods. 
Even on individual categories, our method obtained more than 0.90 on all categories. 

\begin{table}[!]
    \centering
    \scriptsize
    \caption{Number of images per class}
    \begin{tabular}{|c|c|c|c|}
         \hline 
         & Train & Val & Test \\
         \hline
         Authentic & 1,612,202 & 42,382 & 42,397 \\
         \hline
         StarGAN & 28,062 & 738 & 711 \\
         \hline
         CycleGAN & 17,265 & 439 & 439 \\
         \hline 
         ProGAN & 70,286 & 1833 & 1,881 \\
         \hline
         SPADE & 138,075 & 3,717 & 3,704 \\
         \hline 
         StyleGAN & 766,045 & 20,220 & 20,158\\
         \hline 
    \end{tabular}
    \label{tab:n_img_p_class}
\end{table}

\subsection{Tackling newer challenges like StyleGAN2}
Apart from generalization, we tested our method on 100,000 images from the recently released StyleGAN2~\cite{karras2019analyzing} dataset of celebrity faces.
The quality of these images were much better than the previous version and appeared realistic. 
When we tested on this dataset without any fine-tuning, we obtained an accuracy of 0.9464.
This shows that our approach is promising in adapting to newer challenges.
We also fine-tuned to this dataset by adding 100,000 authentic images randomly chosen from different GAN datasets, thus our new dataset comprised of 100,000 authentic images and 100,000 StyleGAN2 images. 
Then, we split this data into 40\% training, 10\% validation and 50\% testing. 
When we trained a new network on this dataset, we obtained a validation accuracy of 0.9984 and testing accuracy of 0.9972, thus also confirming that our approach can be made adjustable to newer GAN datasets.


\subsection{GAN Attribution/Classification}

While the primary area of interest is in determining the authenticity of an image, an immediate extension would be to determine which GAN was used. 
Here we perform an additional experiment on GAN class classification/attribution as a 6-class classification problem, the classes being: Real, StarGAN, CycleGAN, ProGAN, SPADE/GauGAN and StyleGAN.
The number of output layers in the CNN was changed from 1 to 6, and output with the largest value was selected as the estimate.
A breakdown of the number of images per class for training, validation and testing is given in Table~\ref{tab:n_img_p_class}.
First, the network was trained where the input co-occurrence matrices were computed on the whole image.
The training procedure was kept the same as with all other tests in the paper, with the exception of using a batch size of 60, and 10 images from each class per batch.
This encouraged the network to not develop a bias towards any particular GAN for which we have more training data. 
First we consider the images as they are provided in the datasets. 
The classification results are shown in the form of confusion matrices in Table~\ref{tab:conf_no_jpeg-full}.
For convenience, we also report the equal prior accuracy, equal to the average along the diagonal of the confusion matrix.
This equal prior accuracy can be interpreted as the classification accuracy if each class is equally likely.
We obtain an overall classification accuracy (considering equal priors) of 0.9654. 
High classification accuracy was obtained for most categories. 
StyleGAN had comparatively lower accuracy but still more than 90\%, being mostly confused with SPADE/GauGAN and CycleGAN.
These results show that our approach can also be used to identify which category of GAN was used.

Next, we trained the network using a patch size of 128$\times$128 as input, and repeated the experiment.
This is to see how well our method can be used for detection, localization as well as classification. 
The classification results are shown in Table~\ref{tab:conf_no_jpeg}.
Now, we obtain an overall classification accuracy (considering equal priors) of 0.8477 (a drop of 12\% when compared to full image accuracy). 
High classification accuracy was obtained for StarGAN, CycleGAN and ProGAN, while SPADE/GauGAN and StyleGAN had comparatively lower accuracies.
These could be due to many factors such as the number of test images per class, patch size, and the authentic image datasets that were used for training in generating these GAN images. 
In Table~\ref{tab:conf_w_jpeg} we repeat the same experiment (with patch size 128$\times$128) but with images that were randomly preprocessed with JPEG quality factors of 75, 85, 90, or no JPEG compression, with each of the four preprocessing methods equally likely.
For this experiment, the overall classification accuracy drops slightly to 0.8088 due to the impact of JPEG compression.



\begin{figure}[!]
\centering
\includegraphics[width=0.44\textwidth]{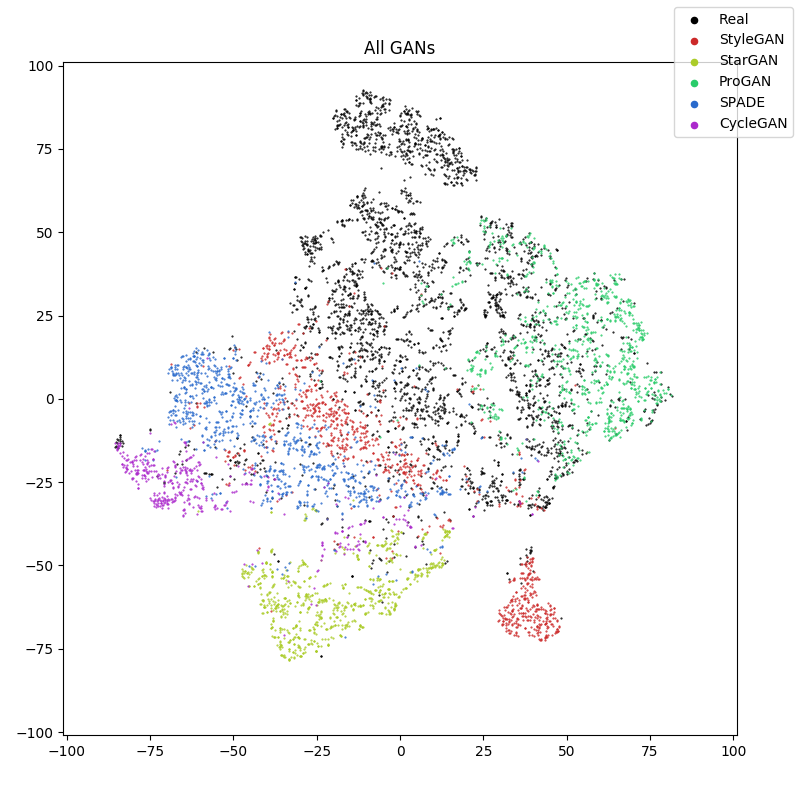}
\vspace{-5pt}
\caption{t-SNE visualization of 6 classes: Real, StyleGAN, StarGAN, ProGAN, SPADE/GauGAN and CycleGAN }
\label{fig:tsne_multi}
\end{figure}

For the multi-class experiment trained without JPEG compression, we repeat the t-SNE visualization procedure. 
Figure~\ref{fig:tsne_multi} shows all data-points on a single plot.
These visualizations further support the results from the classification experiment.

\begin{figure*}[t]
\centering
\subfloat[Real Images]{\includegraphics[width=0.48\textwidth]{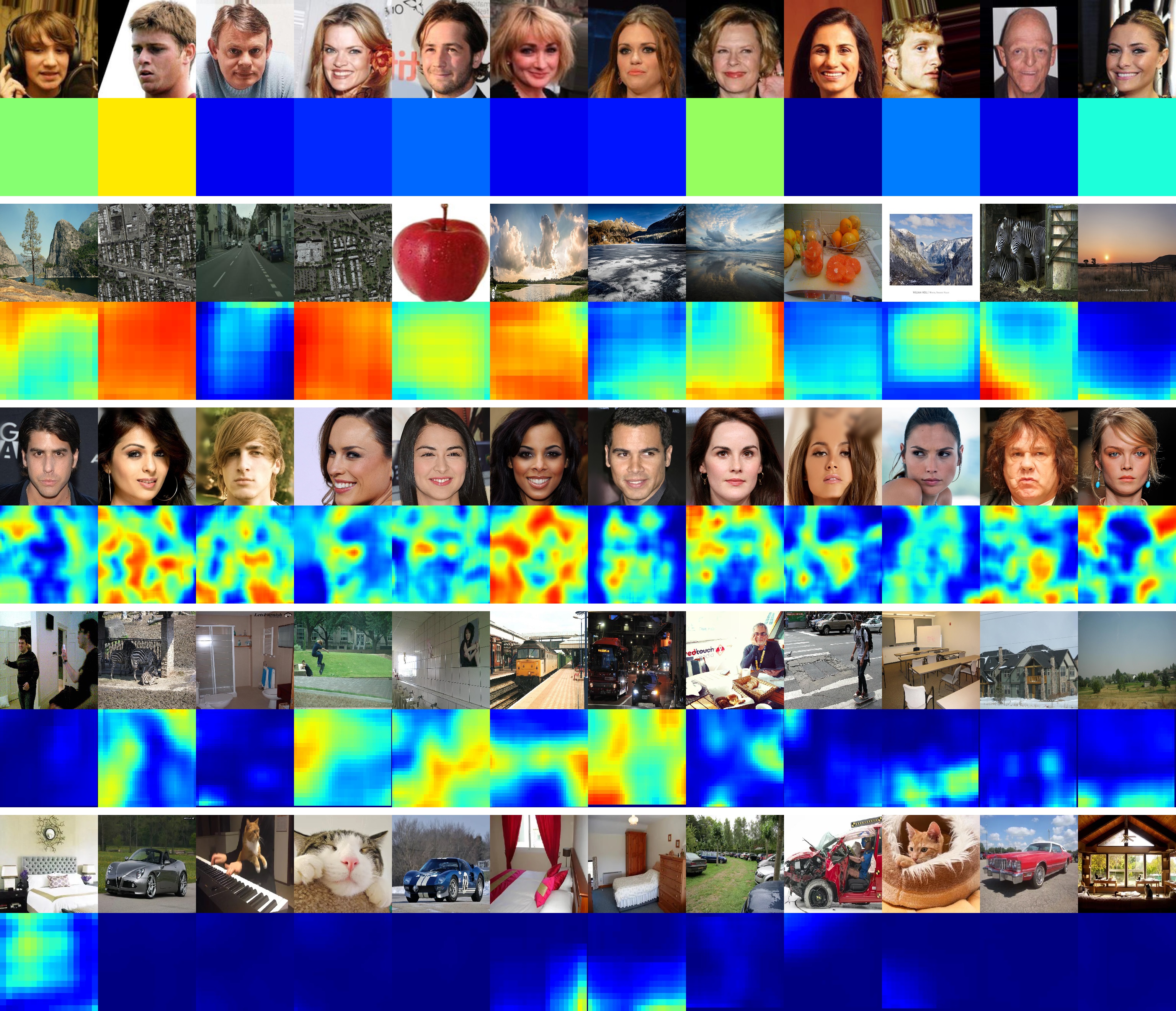}} \ \
\subfloat[GAN Images]{\includegraphics[width=0.48\textwidth]{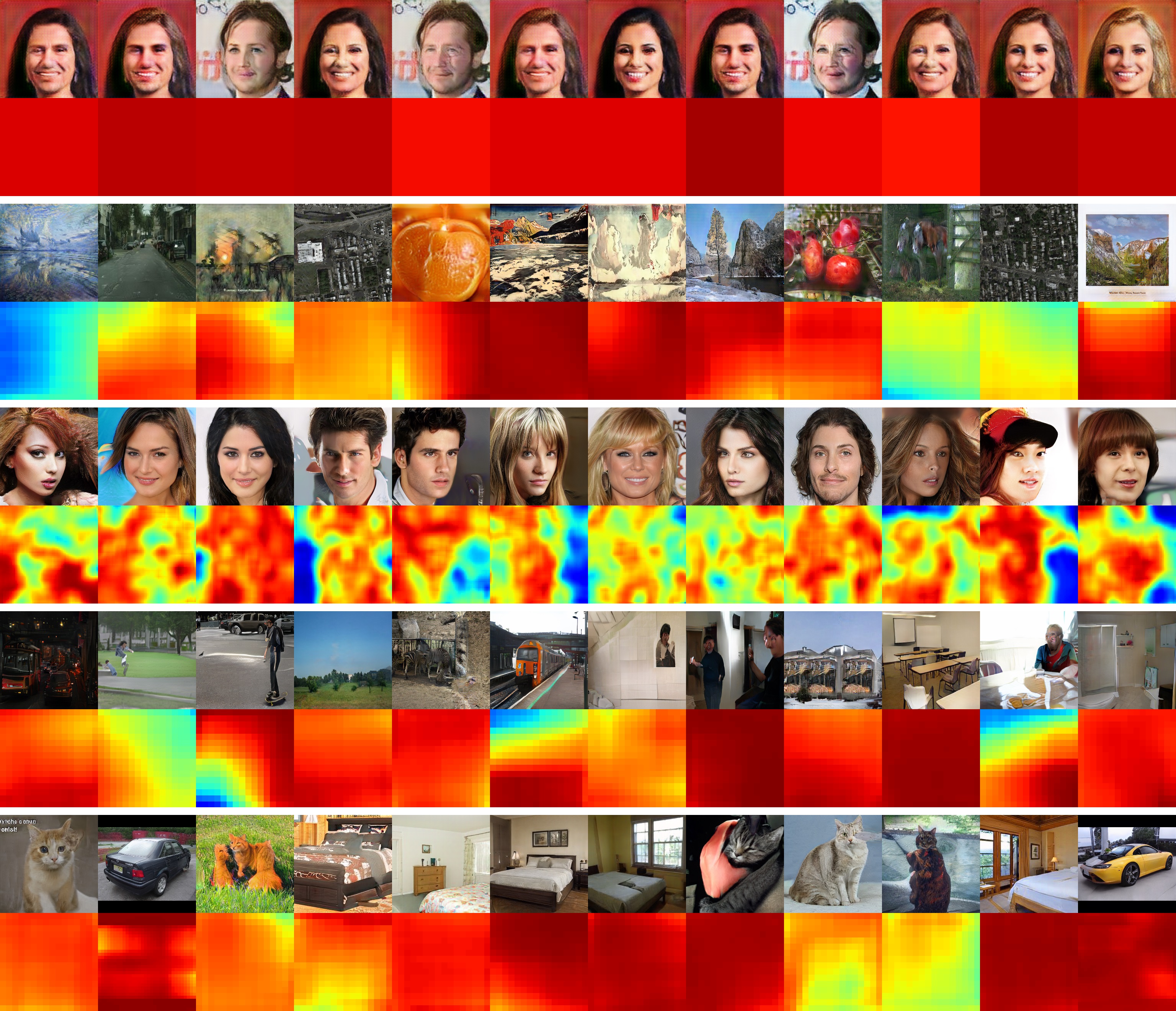}}
\caption{Localization heatmaps of (a) Real images and (b) GAN images from different GAN datasets (top to bottom): ProGAN~\cite{karras2017progressive}, StarGAN~\cite{choi2018stargan}, CycleGAN~\cite{zhu2017unpaired}, StyleGAN~\cite{karras2019style}, and SPADE/GauGAN~\cite{park2019semantic}.}
\label{fig:realgan_heatmaps}
\end{figure*}


\begin{table*}
\caption{Confusion matrix on images from GAN datasets without any pre-processing on the full image. Equal prior accuracy of 0.9654.}
\centering
\begin{tabular}{|c|c|c|c|c|c|c|c|}
\hline
\multicolumn{2}{|c|}{\multirow{2}{*}{}}&\multicolumn{6}{c|}{Predicted Label}\\
\cline{3-8}
\multicolumn{2}{|c|}{} & Real & StarGAN & CycleGAN & ProGAN & SPADE & StyleGAN\\
\hline
\multirow{6}{*}{GT Label} & Real & 0.975\cellcolor{blue!49.6} & 0.000\cellcolor{blue!4.0} & 0.000\cellcolor{blue!5.1} & 0.016\cellcolor{blue!14.4} & 0.002\cellcolor{blue!7.9} & 0.006\cellcolor{blue!10.9}\\
\cline{2-8}
 & StarGAN & 0.000\cellcolor{blue!0.0} & 0.976\cellcolor{blue!49.6} & 0.014\cellcolor{blue!13.9} & 0.000\cellcolor{blue!0.0} & 0.010\cellcolor{blue!12.5} & 0.000\cellcolor{blue!0.0}\\
\cline{2-8}
 & CycleGAN & 0.000\cellcolor{blue!0.0} & 0.000\cellcolor{blue!0.0} & 0.964\cellcolor{blue!49.4} & 0.000\cellcolor{blue!0.0} & 0.036\cellcolor{blue!18.5} & 0.000\cellcolor{blue!0.0}\\
\cline{2-8}
 & ProGAN & 0.000\cellcolor{blue!0.0} & 0.000\cellcolor{blue!0.0} & 0.000\cellcolor{blue!0.0} & 1.000\cellcolor{blue!50.0} & 0.000\cellcolor{blue!0.0} & 0.000\cellcolor{blue!0.0}\\
\cline{2-8}
 & SPADE & 0.001\cellcolor{blue!5.9} & 0.000\cellcolor{blue!4.2} & 0.019\cellcolor{blue!15.3} & 0.000\cellcolor{blue!0.0} & 0.975\cellcolor{blue!49.6} & 0.005\cellcolor{blue!9.9}\\
\cline{2-8}
 & StyleGAN & 0.007\cellcolor{blue!11.3} & 0.000\cellcolor{blue!0.0} & 0.022\cellcolor{blue!16.0} & 0.000\cellcolor{blue!4.4} & 0.068\cellcolor{blue!22.3} & 0.902\cellcolor{blue!48.5}\\
\hline
\end{tabular}
\vspace{-5pt}
\label{tab:conf_no_jpeg-full}
\end{table*}

\begin{table*}
\caption{Confusion matrix on images from GAN datasets without any pre-processing on 128$\times$128 patches. Equal prior accuracy of 0.8477.}
\centering
\begin{tabular}{|c|c|c|c|c|c|c|c|}
\hline
\multicolumn{2}{|c|}{\multirow{2}{*}{}}&\multicolumn{6}{c|}{Predicted Label}\\
\cline{3-8}
\multicolumn{2}{|c|}{} & Real & StarGAN & CycleGAN & ProGAN & SPADE & StyleGAN\\
\hline
\multirow{6}{*}{GT Label} & Real & 0.826\cellcolor{blue!47.2} & 0.003\cellcolor{blue!8.4} & 0.016\cellcolor{blue!14.4} & 0.021\cellcolor{blue!15.8} & 0.066\cellcolor{blue!22.1} & 0.068\cellcolor{blue!22.4}\\
\cline{2-8}
 & StarGAN & 0.000\cellcolor{blue!0.0} & 0.933\cellcolor{blue!49.0} & 0.054\cellcolor{blue!20.9} & 0.000\cellcolor{blue!0.0} & 0.006\cellcolor{blue!10.9} & 0.006\cellcolor{blue!10.9}\\
\cline{2-8}
 & CycleGAN & 0.000\cellcolor{blue!0.0} & 0.002\cellcolor{blue!8.0} & 0.959\cellcolor{blue!49.4} & 0.002\cellcolor{blue!8.0} & 0.032\cellcolor{blue!17.7} & 0.005\cellcolor{blue!9.9}\\
\cline{2-8}
 & ProGAN & 0.000\cellcolor{blue!0.0} & 0.002\cellcolor{blue!7.3} & 0.008\cellcolor{blue!11.6} & 0.981\cellcolor{blue!49.7} & 0.004\cellcolor{blue!9.8} & 0.005\cellcolor{blue!10.2}\\
\cline{2-8}
 & SPADE & 0.001\cellcolor{blue!6.0} & 0.025\cellcolor{blue!16.5} & 0.210\cellcolor{blue!31.3} & 0.008\cellcolor{blue!11.5} & 0.728\cellcolor{blue!45.5} & 0.029\cellcolor{blue!17.3}\\
\cline{2-8}
 & StyleGAN & 0.003\cellcolor{blue!8.7} & 0.025\cellcolor{blue!16.5} & 0.101\cellcolor{blue!25.2} & 0.009\cellcolor{blue!12.0} & 0.203\cellcolor{blue!31.0} & 0.659\cellcolor{blue!44.1}\\
\hline
\end{tabular}
\vspace{-5pt}
\label{tab:conf_no_jpeg}
\end{table*}

\begin{table*}
\caption{Confusion matrix with JPEG compression (128$\times$128 patches). Equal prior accuracy of 0.8088. The images were preprocessed using a JPEG factor of 75, 85, 90, or no compression. Each of these four possible preprocessing functions was randomly selected with equal probability for every image.}
\centering
\begin{tabular}{|c|c|c|c|c|c|c|c|}
\hline
\multicolumn{2}{|c|}{\multirow{2}{*}{}}&\multicolumn{6}{c|}{Predicted Label}\\
\cline{3-8}
\multicolumn{2}{|c|}{} & Real & StarGAN & CycleGAN & ProGAN & SPADE & StyleGAN\\
\hline
\multirow{6}{*}{GT Label} & Real & 0.741\cellcolor{blue!45.7} & 0.005\cellcolor{blue!10.3} & 0.020\cellcolor{blue!15.6} & 0.026\cellcolor{blue!16.8} & 0.103\cellcolor{blue!25.3} & 0.104\cellcolor{blue!25.4}\\
\cline{2-8}
 & StarGAN & 0.006\cellcolor{blue!10.9} & 0.927\cellcolor{blue!48.9} & 0.023\cellcolor{blue!16.2} & 0.000\cellcolor{blue!0.0} & 0.031\cellcolor{blue!17.6} & 0.012\cellcolor{blue!13.4}\\
\cline{2-8}
 & CycleGAN & 0.009\cellcolor{blue!12.2} & 0.014\cellcolor{blue!13.7} & 0.892\cellcolor{blue!48.3} & 0.007\cellcolor{blue!11.2} & 0.074\cellcolor{blue!22.9} & 0.005\cellcolor{blue!9.9}\\
\cline{2-8}
 & ProGAN & 0.002\cellcolor{blue!7.3} & 0.003\cellcolor{blue!8.6} & 0.009\cellcolor{blue!12.1} & 0.973\cellcolor{blue!49.6} & 0.007\cellcolor{blue!11.1} & 0.007\cellcolor{blue!11.1}\\
\cline{2-8}
 & SPADE & 0.075\cellcolor{blue!23.0} & 0.015\cellcolor{blue!14.1} & 0.095\cellcolor{blue!24.6} & 0.009\cellcolor{blue!12.2} & 0.765\cellcolor{blue!46.1} & 0.042\cellcolor{blue!19.3}\\
\cline{2-8}
 & StyleGAN & 0.114\cellcolor{blue!26.1} & 0.021\cellcolor{blue!15.7} & 0.059\cellcolor{blue!21.4} & 0.008\cellcolor{blue!11.5} & 0.243\cellcolor{blue!32.7} & 0.555\cellcolor{blue!41.9}\\
\hline
\end{tabular}
\vspace{-5pt}
\label{tab:conf_w_jpeg}
\end{table*}


\subsection{Localization}
Figure~\ref{fig:realgan_heatmaps} show two example localization outputs.
The image is processed in overlapping patches, with a particular stride and patch size. 
A co-occurrence matrix is then extracted for each patch, and passed through the CNN to produce a score.
For pixels which are a part of multiple patches, the scores are simply the mean of all of the patch responses. These two examples use a patch size of 128, and a stride of 8.
We can see that the heatmaps are predominantly blue for real images and predominantly red for GAN generated images.
This further supports that our method can be effectively used for GAN localization.




\section{Conclusions}

In this paper, we proposed a novel method to detect and attribute GAN generated images, and localize the area of manipulations.
Detailed experimental results using a collection of over 2.7 million GAN and authentic images encompassing 5 major GAN datasets demonstrate that the proposed model is highly effective on a range of image scales and JPEG compression factors. 
In addition, the t-SNE visualization with the neural network deep features showed promising separation of GAN and authentic images using our method.
\begin{acks}
This research was developed with funding from the Defense Advanced Research Projects Agency (DARPA).
The views, opinions and/or findings expressed are those of the author and should not be interpreted as representing the official views or policies of the Department of Defense or the U.S. Government. 
\end{acks}

\bibliographystyle{ACM-Reference-Format}
\bibliography{gan-detection}
\clearpage










\end{document}